\documentclass[10pt,aps,prb,twocolumn,showpacs,amsmath,amssymb,superscriptaddress]{revtex4-1}
\usepackage{graphicx}
\usepackage{natbib}
\usepackage{amsmath}
\usepackage{xcolor}
\usepackage{comment}
\usepackage{soul}
\usepackage[colorlinks=true, urlcolor  = blue, citecolor = blue, linkcolor = blue]{hyperref}

\begin{document}

\title{Magnetic phases of skyrmion-hosting GaV$_4$S$_{8-y}$Se$_{y}$ ($y = 0, 2, 4, 8$) probed with muon spectroscopy}

\author{K\'{e}vin J. A. Franke}
\affiliation{Durham University, Centre for Materials Physics, Durham, DH1 3LE, United Kingdom}

\author{Benjamin M. Huddart}
\affiliation{Durham University, Centre for Materials Physics, Durham, DH1 3LE, United Kingdom}

\author{Thomas J. Hicken}
\affiliation{Durham University, Centre for Materials Physics, Durham, DH1 3LE, United Kingdom}

\author{Fan Xiao}
\affiliation{Laboratory for Neutron Scattering, Paul Scherrer Institut, CH-5232 Villigen PSI, Switzerland}
\affiliation{Department of Chemistry and Biochemistry, University of Bern, CH-3012 Bern, Switzerland}

\author{Stephen J. Blundell}
\affiliation{Oxford University Department of Physics, Clarendon Laboratory, Parks Road, Oxford OX1 3PU, United Kingdom}

\author{Francis L. Pratt}
\affiliation{ISIS Facility, STFC Rutherford Appleton Laboratory, Chilton, Didcot, Oxfordshire, OX11 0QX, United Kingdom}

\author{Marta Crisanti}
\affiliation{University of Warwick, Department of Physics, Coventry, CV4 7AL, United Kingdom}
\affiliation{Institut Laue-Langevin, CS 20156, 38042 Grenoble Cedex 9, France}

\author{Joel A. T. Barker}
\affiliation{Laboratory for Muon Spin Spectroscopy, Paul Scherrer Institut, CH-5232 Villigen PSI, Switzerland}

\author{Stewart J. Clark}
\affiliation{Durham University, Centre for Materials Physics, Durham, DH1 3LE, United Kingdom}

\author{Ale\v s \v Stefan\v ci\v c}
\affiliation{University of Warwick, Department of Physics, Coventry, CV4 7AL, United Kingdom}

\author{Monica Ciomaga Hatnean}
\affiliation{University of Warwick, Department of Physics, Coventry, CV4 7AL, United Kingdom}

\author{Geetha Balakrishnan}
\affiliation{University of Warwick, Department of Physics, Coventry, CV4 7AL, United Kingdom}

\author{Tom Lancaster}
\affiliation{Durham University, Centre for Materials Physics, Durham, DH1 3LE, United Kingdom}

\date{\today}

\begin{abstract}
We present the results of a muon-spin spectroscopy investigation of GaV$_4$S$_{8-y}$Se$_{y}$ with $y=0, 2, 4$ and 8. 
Zero-field measurements suggest that GaV$_{4}$Se$_{8}$ and GaV$_{4}$S$_{8}$ have distinct magnetic ground states, with the latter material showing an anomalous temperature-dependence of the local magnetic field. It is not possible to evolve the magnetic state continuously between these two systems, with the intermediate $y=2$ and $4$ materials showing glassy magnetic behaviour at low temperature. The skyrmion lattice (SkL) phase is evident in the $y=0$ and 8 materials through an enhanced response of the muon-spin relaxation to the emergent dynamics that accompany the SkL. 
For our polycrystalline samples of GaV$_4$Se$_{8}$, this enhanced dynamic response is confined to a smaller region of the magnetic field-temperature phase diagram than the previous reports of the SkL in single crystals.
\end{abstract}
\maketitle

\section{Introduction}
\label{intro}
In recent years, a number of spectacular advances have demonstrated the existence, not only of magnetic skyrmions, but also their ordering into a skyrmion lattice (SkL). The potential of skyrmions as high-density, low-power, data-storage and logic devices is driving their exploration.\cite{fert_skyrmions_2013,zhang_topological_2015} Skyrmions are topological spin textures stabilized by the competition between the antisymmetric Dzyaloshinskii-Moriya interaction (DMI)\cite{dzyaloshinskii_thermodynamic_1958, moriya_anisotropic_1960} and the symmetric exchange interaction. 
 The emergence of a Bloch-type SkL phase has been observed in several magnetic materials with a chiral structure close to the paramagnetic (PM) to helical magnetically-ordered phase boundary. These include the B20 compounds,\cite{muhlbauer_skyrmion_2009,yu_real-space_2010, seki_observation_2012} and $\beta$-Mn-type CoZnMn alloys.\cite{tokunaga_new_2015} In ultra-thin magnetic layers \cite{heinze_spontaneous_2011,hsu_electric-field-driven_2017} and multilayers\cite{chen_room_2015,moreau-luchaire_additive_2016} where inversion symmetry is broken at interfaces,  N\'eel-type skyrmions can be stabilized. The formation of N\'eel-type skyrmions in bulk crystals with polar $C_{nv}$ symmetry was first predicted by Bogdanov and Yablonskii,\cite{bogdanov_thermodynamically_1989} and was recently observed in VOSe$_2$O$_5$,\cite{kurumaji_neel-type_2017} GaV$_4$S$_8$,\cite{kezsmarki_neel-type_2015,white_direct_2018} and GaV$_4$Se$_8$.\cite{bordacs_equilibrium_2017}

The crystal structure of GaV$_4$S$_8$ and GaV$_4$Se$_8$ is composed of (GaX$_4$)$^{5-}$ tetrahedra and (V$_4$X$_4$)$^{5+}$ distorted heterocubane units (where  $X=$ S, Se), organized in a rock salt-type arrangement (space group $F\bar{4}3m$). 
\cite{pocha_electronic_2000} Magnetic moments result from one unpaired electron per metallically-bonded V${_4}$ tetrahedron, causing the V$_{4}$ units to carry an effective spin $S = 1/2$.\cite{pocha_electronic_2000,bordacs_equilibrium_2017, zhang_magnetic_2017, johrendt_crystal_1998}  [Both compounds are insulators due to a large distance ($\approx 4$ \AA) between V${_4}$ clusters.] At room temperature, these materials have a non-centrosymmetric cubic crystal structure with $T_{d}$ point symmetry, but at around 40~K (in both materials) a Jahn-Teller transition 
changes this to a rhombohedral polar $C_{3v}$ symmetry by stretching the lattice along one of the four $\langle 111 \rangle$ cubic axes.\cite{kezsmarki_neel-type_2015,butykai_characteristics_2017,bordacs_equilibrium_2017} Below the respective magnetic ordering temperatures ($T_{\mathrm{c}}\approx 12.5$~K for \textrm{GaV$_4$S$_{8}$};\cite{nakamura_low-field_2009,kezsmarki_neel-type_2015}  and $T_{\mathrm{c}}\approx 17.5$~K for \textrm{GaV$_4$Se$_{8}$} \cite{fujima_thermodynamically_2017,bordacs_equilibrium_2017}), applied magnetic fields drive successive transitions between cycloidal (C), SkL, and field-polarized (FP) phases. In contrast to the Bloch-type SkL plane commonly observed to align perpendicular to the applied magnetic field direction, the lattice vectors of the N\'eel-type SkL are constrained to a plane perpendicular to the rhombohedral lattice distortion.\cite{kezsmarki_neel-type_2015,ehlers_skyrmion_2016, leonov_skyrmion_2017, leonov_asymmetric_2017} 
For GaV$_4$S$_8$, a uniaxial magnetocrystalline anisotropy is found along this distortion, thought to favor ferromagnetic (FM) alignment below $T_{\mathrm{FM}} \approx 5$~K.\cite{kezsmarki_neel-type_2015,white_direct_2018} 
In contrast to other bulk SkL-hosting systems, the SkL in GaV$_{4}$S$_{8}$ was observed over an increased temperature region, persisting from $T_{\textrm{c}}$ down to $9$~K ($\approx 0.7$ $T_{\textrm{c}}$). 
In GaV$_4$Se$_8$ the SkL was reported to extend down to zero Kelvin. The large temperature range over which the SkL is observed in both of these compounds,  and the fact that substituting S for Se appears to further increase the stability of the SkL, makes intermediate compositions in the GaV$_4$S$_{8-y}$Se$_{y}$ series interesting to investigate, in an attempt to continuously tune the stability region of the SkL between both end compounds, or even find a further increase in the extent of the SkL region in the phase diagram. 

The use of transverse field (TF) muon-spin rotation ($\mu^{+}$SR) 
to investigate the SkL was motivated by its use in probing the vortex lattice in type-II superconductors, where the technique provides measurements of the internal field distribution caused by the magnetic field texture. 
It has been used to probe the SkL region in bulk Cu$_2$OSeO$_3$,\cite{lancaster_transverse_2015} and thin films of MnSi\cite{lancaster_transverse_2016} and FeGe.\cite{zhang_room-temperature_2017}
In contrast, the use of longitudinal field (LF) $\mu^{+}$SR on skyrmion-hosting materials has not been reported in detail, despite initial hints that it was effective at probing the emergent dynamics that accompany the skyrmion phase.\cite{liu_sr_2016} Here, we present the results of a $\mu^{+}$SR investigation of skyrmion-hosting materials GaV$_4$S$_8$ and GaV$_4$Se$_8$ and intermediate compounds from the GaV$_4$S$_{8-y}$Se$_{y}$ series, with $y=2$ and 4. 
We present measurements in the LF, TF, and zero-field (ZF) geometries, allowing access to static and dynamic local magnetic properties.
While GaV$_4$S$_8$ and GaV$_4$Se$_8$ are very similar in exhibiting a SkL phase, we show below that their magnetic behaviour is quite different. This is likely attributable to subtle (but significant) differences in the electronic ground states of the two systems. We find that the intermediate compounds with $y=2$ and $4$ do not undergo a transition to a state of long-range magnetic order, but instead show a glassy freezing of dynamics as the temperature is reduced.  
Our LF $\mu^{+}$SR measurements reveal that the skyrmion phases in GaV$_4$S$_8$ and GaV$_4$Se$_8$ give rise to emergent dynamics on the muon (microsecond) time scale, reflecting the slowing of magnetic fluctuations perpendicular to the applied field. In GaV$_4$S$_8$ these dynamics are observed over the temperature and field range where the SkL was observed in single crystals.\cite{kezsmarki_neel-type_2015,white_direct_2018} For GaV$_4$Se$_8$, the observed dynamics are more limited in their extent in the $H$--$T$ phase diagram than the reported extent of the SkL.\cite{bordacs_equilibrium_2017} It is thus plausible that the SkL region of GaV$_4$Se$_8$ is less extensive than previously suggested. 

This paper is structured as follows: in Section~\ref{sec:expt} we describe our methods; in Section~\ref{sec:susc} we discuss the characterization of our samples using AC and DC magnetometry; in Section~\ref{sec:zfmu} we describe ZF measurements on the $y=0, 2, 4$ and 8 materials; 
in Section~\ref{sec:lfmu} we turn to the dynamics of the materials and reveal the dynamics signature of the SkL seen in LF measurements, while in Section~\ref{sec:tfmu} we present TF measurements investigating static local field distributions.
In Section~\ref{sec:musite} we present the results of calculations of the nature of the muon sites in this system. 
We discuss our findings and their consequences in Section~\ref{sec:discussion} and finally present our conclusions in Section~\ref{sec:conclusions}.

\section{Experimental}
\label{sec:expt}
Polycrystalline samples of GaV$_4$S$_{8-y}$Se$_{y}$ ($y=0,2,4,8$) were prepared by reacting stoichiometric amounts of high-purity elements (Ga, V, S and Se) in evacuated silica tubes. The samples were heated at a rate of $10$ $^{\circ}$C/h to $810$--$830$ $^{\circ}$C, kept at this temperature for $300$ h, and afterwards water quenched. 

DC magnetization measurements were performed using a Quantum Design MPMS. 
AC susceptibility measurements were performed on the same instrument at an excitation frequency of $10$~Hz with an amplitude of $0.1$~mT.

In a ZF $\mu^{+}$SR measurement,\cite{blundell_spin-polarized_1999, reotier_muon_1997} spin-polarized muons are implanted in a magnetic material. Muons stop at random positions on the length scale of the field texture where they precess about the total local magnetic field $B$ at the muon site, with frequency $\omega = \gamma_{\mu}B$, where $\gamma_{\mu} = 2 \pi \times 135.5$~MHz T$^{-1}$ is the muon gyromagnetic ratio. 
The observed property of the experiment is the time evolution of the asymmetry $A(t)$, which is proportional to the average muon spin polarization $P_{z}(t)$.
In a TF $\mu^{+}$SR experiment, \cite{blundell_spin-polarized_1999, reotier_muon_1997} a magnetic field $H$ is applied perpendicular to the initial muon spin direction.
In a LF $\mu^{+}$SR experiment, the external field $H$ is applied in the direction of the initial muon-spin polarization, suppressing the contribution from static magnetic fields at the muon site. This allows us to probe the dynamics of the system, as time-varying magnetic fields at the muon site are able to flip muon spins and therefore to relax the average muon polarization. 

We performed ZF and TF $\mu^{+}$SR measurements using the General Purpose Surface-Muon Instrument (GPS) at S$\mu$S. LF and ZF measurements were carried out using the HiFi spectrometer at the ISIS muon source. Polycrystalline samples were packed into Ag foil packets (foil thickness 25~$\mu$m) covering approximately $2 \times 2$ cm$^2$ in area and mounted in a He4 cryostat. For measurements on HiFi, the sample was mounted on a silver backing plate, while on GPS the packet was suspended in the beam on a fork. Data analysis was carried out using the WiMDA analysis program.\cite{PRATT2000710} For all measurements, the sample was warmed above the magnetic ordering temperature $T_{\mathrm{c}}$ and cooled in a fixed applied magnetic field $H$. Measurements were made on warming.

To understand the differences in the response of the muon in GaV$_{4}$S$_{8}$ and GaV$_{4}$Se$_{8}$ we carried out density functional theory (DFT) calculations using the plane-wave basis set electronic structure code \textsc{castep}.\cite{CASTEP, monkhorst_special_1976, lejaeghere_reproducibility_2016, moeller} In order to identify muon stopping sites in this material, we carried out spin-polarized DFT calculations with the generalized gradient approximation (GGA).\cite{PBE} Further details of these calculations can be found in the SI.\cite{SupMat}

\section{Magnetization measurements}
\label{sec:susc}
\begin{figure}
\includegraphics{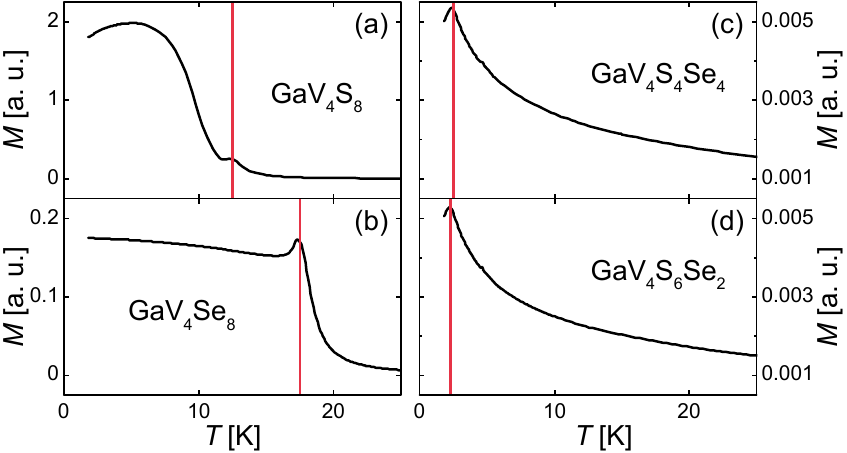}
\caption{\label{DC-10mT}Magnetization $M$ of (a) GaV$_4$S$_8$, (b) GaV$_4$Se$_8$, (c) GaV$_4$S$_4$Se$_4$, and (d) GaV$_4$S$_6$Se$_2$ as a function of temperature in an applied magnetic field of $10$ mT. Vertical lines indicate the location of the transition from a magnetically ordered to the paramagnetic phase.}
\end{figure}

Magnetization measurements made in an applied field of $10$~mT as a function of increasing temperature are shown in Fig.~\ref{DC-10mT}. The transition from a magnetically ordered to the paramagnetic phase is observed as a peak in the magnetization and indicated by a vertical line.\cite{ruff_multiferroicity_2015}  While the transition is observed at $T_{\mathrm{c}}=12.5$ for GaV$_4$S$_8$ and $T_{\mathrm{c}}=17.5$ K for GaV$_4$Se$_8$, it drops to $T_{\mathrm{c}}=2.3$ and $T_{\mathrm{c}}=2.5$ K for GaV$_4$S$_6$Se$_2$ and GaV$_4$S$_4$Se$_4$, respectively. This constitutes a dramatic decrease in the critical temperature for the intermediate compositions in the GaV$_4$S$_{8-y}$Se$_{y}$ series.

\begin{figure}[b]
\includegraphics[width=\columnwidth]{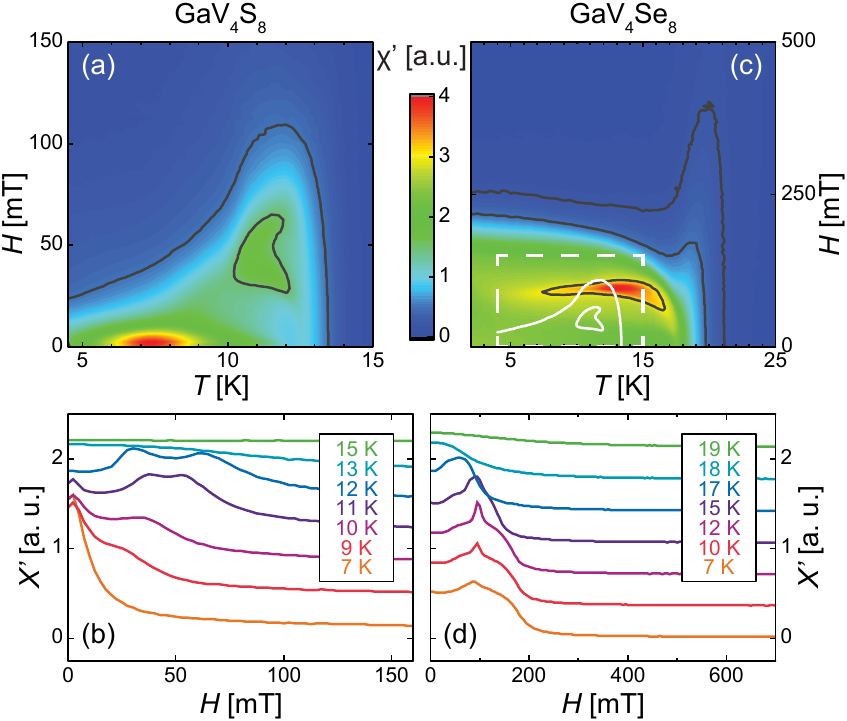}
\caption{\label{ACsusc}{\it Top}: Mapping of the phase diagrams of (a) GaV$_4$S$_8$ and (c) GaV$_4$Se$_8$ using the real part $\chi'$ of the AC susceptibility. White lines show the GaV$_{4}$S$_{8}$ phase diagram for comparison. {\it Bottom}: Real part $\chi'$ of the AC susceptibility as a function of applied field for selected temperatures for (b) GaV$_4$S$_8$ and (d) GaV$_4$Se$_8$, respectively. Curves have been offet for clarity.} 
\end{figure}

Fig.~\ref{ACsusc}(a) shows the real part $\chi'$ of the AC susceptibility from measurements on GaV$_4$S$_8$. Whereas $\chi'$ is commonly observed to decrease in the SkL phase relative to the conical phase in materials such as B20 compounds and Cu$_2$OSeO$_3$,\cite{seki_observation_2012,bauer_magnetic_2012,bauer_generic_2016} we see an increase in $\chi'$ in the region of the phase diagram where the SkL phase has been observed.\cite{kezsmarki_neel-type_2015} This suggests that the magnetization in this system more closely follows the excitation field in the SkL than the surrounding phases. Broad susceptibility peaks [Fig.~\ref{ACsusc}(b)] indicate slow dynamics at the boundaries between C, SkL, and FP phases. The same increase in $\chi'$ surrounded by peaks in the susceptibility were reported for GaV$_4$S$_8$ single crystals by Butykai {\it et al.}\cite{butykai_relaxation_2017} As a consequence, we attribute the enhancement of susceptibility coinciding with the SkL phase directly to the SkL and not to an angular average over peaks marking phase transitions. As noted previously,\cite{kezsmarki_neel-type_2015} the SkL phase is located between the C, FM, and PM phases, which is different from typical skyrmion hosting materials, where the SkL occurs in a small pocket embedded in the helical part of the phase diagram close to $T_{\mathrm{c}}$. 

For GaV$_4$Se$_8$ a region of increased $\chi'$ is observed near $T_{\mathrm{c}}$, in a part of the phase diagram more consistent with the usual location of the SkL [Fig.~\ref{ACsusc}(c)]. 
The extent of this phase in the $H$--$T$ diagram is significantly smaller than previously reported for single crystal samples.\cite{fujima_thermodynamically_2017, bordacs_equilibrium_2017}
For single crystals, a peak in $\chi'$ at around $500$ mT was interpreted as the signature of a transition between the SL and FP phase. Such a peak is not observed in our data collected on polycrystalline samples [Fig.~\ref{ACsusc}(d)].
Fig.~\ref{ACsusc}(c) compares the phase diagrams of the two systems, demonstrating that in the Se-containing system the region of enhanced susceptibility extends to higher temperature and field and thus is characterized by a larger energy scale.

\section{ZF $\mu^{+}$SR measurements}
\label{sec:zfmu}
\begin{figure}
\includegraphics{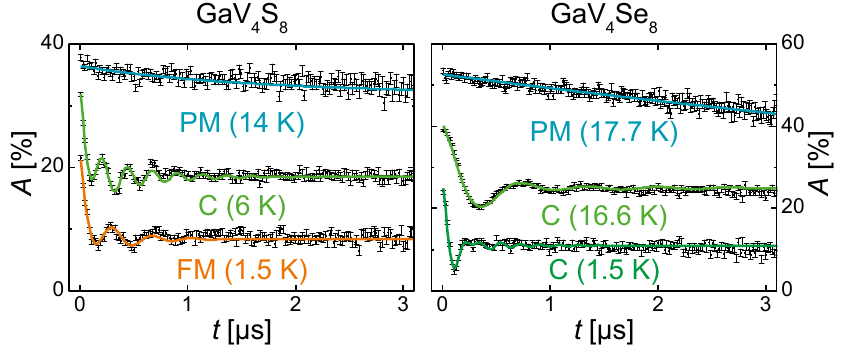}
\caption{\label{TDspectra} Typical time-domain spectra measured for GaV$_4$S$_8$ (left) and GaV$_4$Se$_8$ (right) in ZF in the FM, C, and PM phases. Lines are fits described in the text and the curves have been offset vertically for clarity.}
\end{figure}

Example time-domain spectra from ZF $\mu^{+}$SR measurements are shown in Fig.~\ref{TDspectra}. Oscillations are observed for measurements in the FM and C phases, characteristic of the presence of quasi-static long-range magnetic order. Above $T_{\mathrm{c}}$, in the PM region, the oscillations vanish and the asymmetry $A(t)$ relaxes only weakly relative to the relaxation of the signal in the LRO regime.  

To parametrize the behaviour of both GaV$_4$S$_8$ and GaV$_4$Se$_8$, the spectra were fitted to a function
\begin{equation*}
\begin{split}
A(t)&=  A_{1}{\rm e}^{-\lambda_{1} t} \textrm{cos}(\gamma_{\mu} B_{1} t+\psi)+A_{2}{\rm e}^{-\lambda_{2} t} \textrm{cos}(\gamma_{\mu} B_{2}  t) \\
& +A_{3}{\rm e}^{-\lambda_{3} t}+A_{\mathrm{bg}},
\end{split}
\end{equation*}
where $A_{\mathrm{bg}}$ accounts for those muons that stop in the sample holder or cryostat tails. The third component exhibits a constant amplitude $A_{3}$ and relaxation rate $\lambda_{3} \ll \lambda_{1,2}$ throughout the investigated temperature ranges. 

\begin{figure}
\includegraphics{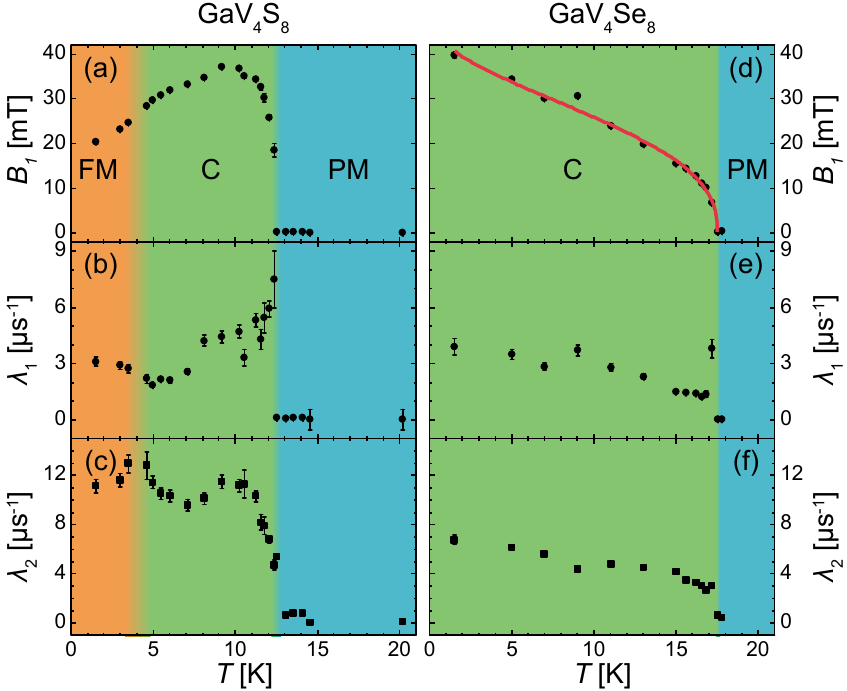}
\caption{\label{ZFmuSR} Results of fitting ZF data for GaV$_4$S$_8$ (a-c) and GaV$_4$Se$_8$ (d-f). Evolution of the internal field (a \& d) and the relaxation rates $\lambda_{1}$ (b \& e) and $\lambda_{2}$ (c \& f) with $T$.}
\end{figure}

The observation of two oscillatory components corresponds to the occurrence of two magnetically distinct muon sites in each material.
Best fits were obtained with the second field component constrained such that $B_{2} = a B_{1}$, with scaling constant $a_{\textrm{GaV$_4$S$_8$}}=0.25$ and $a_{\textrm{GaV$_4$Se$_8$}}=0.66$ for all temperatures. Fitted parameters are plotted in Fig.~\ref{ZFmuSR}. 

For GaV$_4$Se$_8$ the internal field shows the expected decrease with increasing temperature [Fig.~\ref{ZFmuSR}(d)]. The transition from C to the PM phase is observed as a sharp peak in $\lambda_{1}$ and as an overall drop in the magnitude of both relaxation rates [Figs~\ref{ZFmuSR}(e) and (f)]. However, the behaviour for GaV$_4$S$_8$ is more complex. 
We observe the expected decrease in the internal field with increasing temperature only above $T=10$~K [Fig.~\ref{ZFmuSR}(a)]. However, below $10$~K the internal field is found to {\it increase} approximately linearly with increasing temperature. This unusual trend does not seem to be affected by the transition from FM to C phase that can be observed as a peak in the relaxation rate $\lambda_{2}$ [Fig.~\ref{ZFmuSR}(c)]. The form of the oscillations is observed to change in the FM phase, with the phase $\psi$ taking different values in the FM ($53^{\circ}$) and C ($31^{\circ}$) phases. Note, that our data cannot be fitted with a Bessel function (which is often approximated by a damped cosine with $\psi=-45^{\circ}$), as might be expected for uniform sampling of an incommensurate magnetic texture.\cite{reotier_muon_1997} The need for a phase factor is unusual for a uniform magnetization expected in the FM region, and in conjunction with the increase in $B_{1}$ with temperature, is suggestive of the magnetic state in this system being more complicated than the simple ferromagnetic structure previously assumed. We return to this in Section~\ref{sec:discussion}.

\begin{figure}
\includegraphics{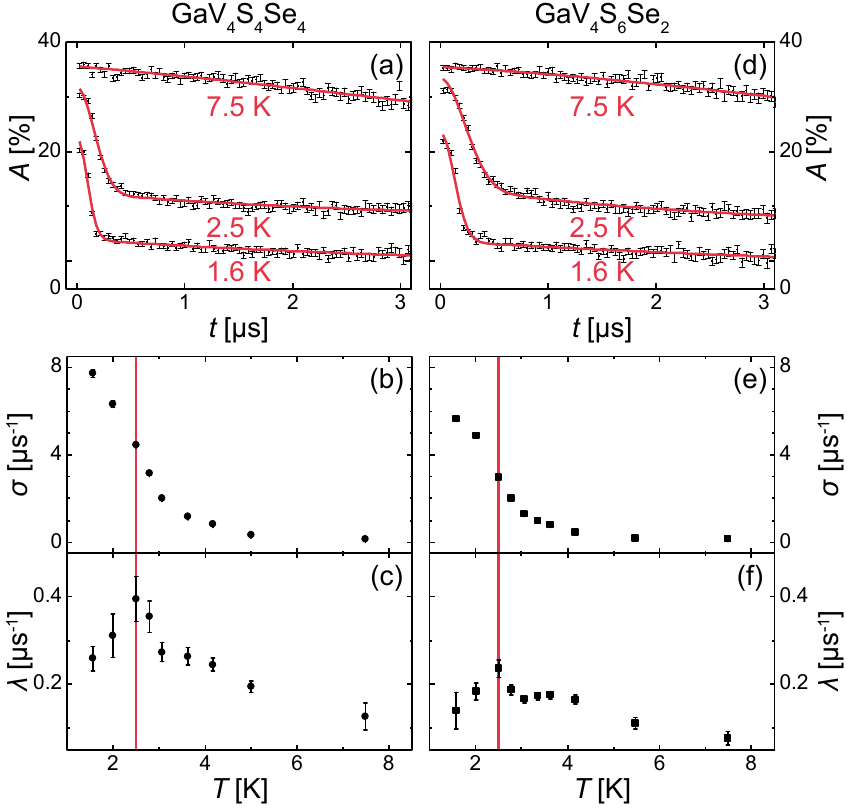}
\caption{\label{ZFmuSRdoped}Example ZF time domain spectra, and results of fitting for (a) - (c) GaV$_4$S$_4$Se$_4$, and (d) - (f) GaV$_4$S$_6$Se$_2$.}
\end{figure}

To elucidate the difference in magnetic properties between GaV$_4$S$_8$ and GaV$_4$Se$_8$, and investigate the possibility to  tune magnetic properties in the GaV$_4$S$_{8-y}$Se$_{y}$ series, we made ZF measurements on the intermediate compounds GaV$_4$S$_4$Se$_4$ ($y=4$) and GaV$_4$S$_6$Se$_2$ ($y=2$). Neither of these compounds exhibits oscillations in the asymmetry [Figs~\ref{ZFmuSRdoped}(a) \& (d)]. Instead, the spectra show monotonic relaxation at all temperatures, found to be best fitted by the function
\begin{equation*}
A(t)=  A_{1}{\rm e}^{-(\sigma t)^{2}} +A_{2}{\rm e}^{-\lambda t} +A_{\mathrm{bg}}.
\end{equation*}
Note that it is not possible to fit the data with the product of a Gaussian and an exponential relaxation, which would model the separation between static and dynamic contributions at a single muon site.
The need for two components thus suggests the presence of two distinct environments for muons stopped in these materials. Typically, a Gaussian relaxation approximately describes relaxation due to a static array of disordered spins (with some residual dynamics). 
If the spin configuration results in a normal internal field distribution, we would expect such relaxation with $\sigma =\gamma_{\mu} \sqrt{\langle (B-B_{0})^{2}\rangle}$. 

In contrast, exponential relaxation reflects dynamic fluctuations in the magnetic field distribution. In a system of dense local moments in the fast fluctuation limit, dynamics will lead to relaxation at a rate in ZF given approximately by $\lambda = 2\gamma_{\mu}^{2}\Delta^{2}\tau$, where $\tau$ is the correlation time and $\Delta$ the width of the internal field distribution. 
The addition of these two components suggests that a fraction of the muons experience static disorder, while the remainder experience dynamic field fluctuations. 
In both materials a gradual decrease in $\sigma$ with increasing temperature is suggestive of a magnetic field distribution whose width grows as the temperature is reduced [Fig.~\ref{ZFmuSRdoped}(b) \&(e)]. A peak in the exponential relaxation rate $\lambda$ around $T=2.5$ K [Fig.~\ref{ZFmuSRdoped}(c) \& (f)] coincides with the critical temperature established in DC magnetization measurements, but is not reflected by any marked change in $\sigma$. It is notable that this characteristic temperature is significantly lower than the ordering temperatures of $T_{\mathrm{c}}=12.5$~K and $T_{\mathrm{c}}=17.5$~K observed for GaV$_4$S$_8$ and GaV$_4$Se$_8$, respectively.
These narrow peaks in $\lambda$ are suggestive of a sudden freezing out of dynamics below 2.5~K. We reason that, upon cooling, dynamics slow down and pass through the window of excitation frequencies probed by $\mu^{+}$SR (MHz--GHz). This assessment is supported by the observation of a peak in the ratio $A_{2}/A_{1}$ close to this characteristic temperature indicating a dominant contribution of dynamic fluctuations on the muon time scale. However, it is likely that the material retains a degree of disorder down to the lowest temperatures. This provides a picture of a glassy freezing out of spins and at least partially ordered ground state for these two materials. 
Our results show that although GaV$_4$S$_8$ and GaV$_4$Se$_8$ are structurally very similar and both exhibit a SkL phase, they are magnetically very different. Moreover, their magnetic phase diagrams cannot be continuously transformed from one into the other using random substitutions of S with Se, since the disorder introduced leads to a glassy ground state. This behaviour is reminiscent of that shown by GaV$_{4-x}$Mo$_{x}$S$_{8}$ ($0\leq x \leq 4$), where Mo is substituted for V. \cite{powell2007}

\section{LF $\mu^{+}$SR measurements}
\label{sec:lfmu}
To probe dynamics across the phase diagram LF measurements were made on GaV$_4$S$_8$ and GaV$_4$Se$_8$. In polycrystalline samples we expect $2/3$ of the muon spin components to lie perpendicular to the local field and give rise to oscillations whereas the remaining $1/3$ align along the direction of the local magnetic fields. We utilize the long time window at ISIS to measure slow relaxation in the $1/3$ tail. This asymmetry can only be relaxed by dynamic fluctuations of the magnetic field distribution at the muon site, which can cause transitions between the muon spin--up and --down states, split in energy by the applied magnetic field. 
\begin{figure}
\includegraphics{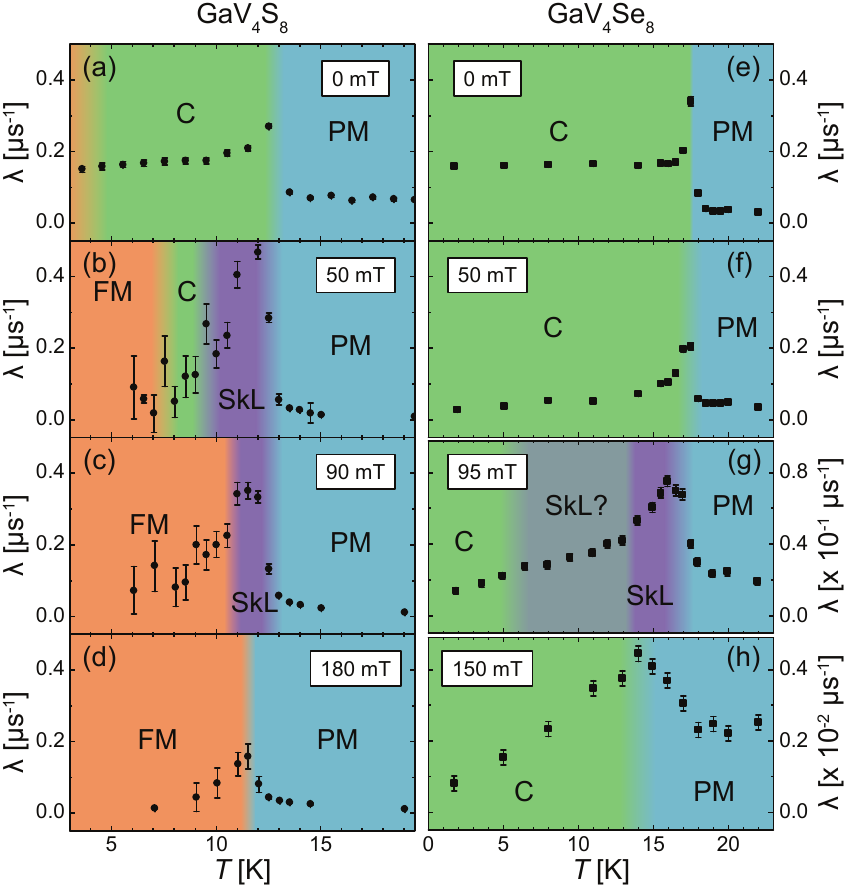}
\caption{\label{LFmuSR} (a) and (e) Results of fitting ZF data for GaV$_4$S$_8$ and GaV$_4$Se$_8$. Phases determined using Fig.~\ref {ZFmuSR}. (b)-(d) and (f)-(h) Fits of LF data.}
\end{figure}

ZF and LF $\mu^{+}$SR time domain spectra measured at ISIS were found to decay following a simple exponential relaxation and the asymmetry was thus fitted to a function of the form
\begin{equation*}
A(t)=A\,e^{-\lambda t}+A_{\mathrm{bg}}.
\end{equation*}

Taking the applied field to define the $z$ direction, the relaxation rate in the fast fluctuation limit is approximated by $\lambda = \gamma_{\mu}^{2}\langle \Delta_{x}^{2} +\Delta_{y}^{2} \rangle \tau$, where $\tau$ is the correlation time for fluctuations in the distribution $\Delta_{x,y}$ of magnetic field components in both the $x$ and $y$ directions (expected to be equivalent in our polycrystalline samples). 
Results from fitting ZF and LF data are presented in Fig.~\ref{LFmuSR}. ZF data on GaV$_4$S$_8$ and GaV$_4$Se$_8$ show a narrow peak in $\lambda$ at the transition between the PM and C phases. 
For GaV$_4$S$_8$, the peak in $\lambda$ that marks the transition between magnetically ordered and disordered phases is larger and significantly broadened for applied fields of $50$~mT and $90$~mT. The temperature and field region of this sizeable feature coincides with where the SkL phase has been observed previously \cite{kezsmarki_neel-type_2015} and is consistent with our AC susceptibility measurements  (Fig.~\ref{ACsusc}) and our TF $\mu^{+}$SR. We conclude that the SkL can be detected via a sizeable contribution to the relaxation rate in LF $\mu^{+}$SR measurements. 

For measurements on  GaV$_4$Se$_8$, the transition between the magnetically ordered and disordered phases is again seen as a narrow peak in the relaxation rates. However, we do not observe a significant qualitative change in dynamics in the magnetically ordered phase for measurements made with $\mu_{0}H=150$~mT where the SkL was previously observed. Instead, a roughly linear increase of $\lambda$ is observed with temperature as is the case for GaV$_4$S$_8$ at $180$~mT. (Above the phase transition at $T_{\mathrm{c}}=14$~K,  a very gradual decrease in the relaxation rate suggesting an intermediate region of behaviour, c.f.~our TF data.)
However, for measurements made at $\mu_{0}H=95$~mT, we  observe the characteristic enhancement of magnitude and broadened peak in $\lambda$, similar to that observed in the SkL region of GaV$_{4}$S$_{8}$ [Fig.~\ref{LFmuSR}(g)]. A slightly increased $\lambda$ can even be observed down to $6.5$~K below which the relaxation rate decreases. Following our interpretation of the results obtained on GaV$_4$S$_8$, we ascribe this increase in $\lambda$ to the presence of the SkL phase. 

\section{TF $\mu^{+}$SR measurements}
\label{sec:tfmu}

\begin{figure}
\includegraphics{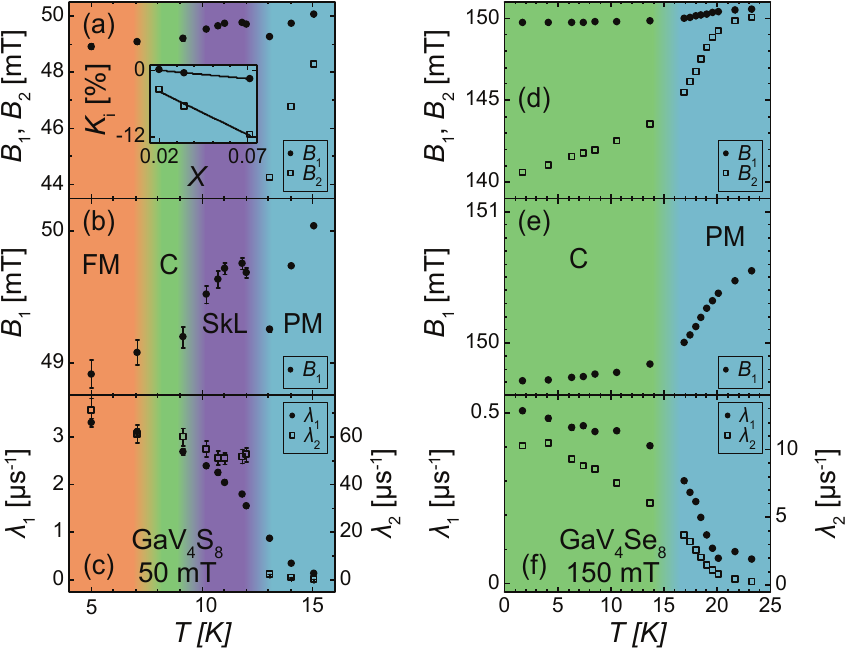}
\caption{\label{TFmuSR} {\it Left}: Results of fitting TF data on GaV$_4$S$_8$ in $50$~mT. (a) Internal fields $B_{1}$ and $B_{2}$.  Inset: muon Knight shifts $K_{i}$. (b) Enlarged view of $B_{1}$ showing an increase in the SkL. (c) Relaxation rates $\lambda_{1}$ and $\lambda_{2}$. (Location of the C phase according to LF measurements is indicated.) 
{\it Right}: Fitting of TF data of GaV$_4$Se$_8$ in $150$~mT. (g) Internal fields $B_{1}$ and $B_{2}$. (h) Enlarged view of $B_{1}$. (i) Relaxation rates $\lambda_{1}$ and $\lambda_{2}$.}
\end{figure}

To investigate the local field distributions across the phase diagrams of the $y=0$ and 8 materials, TF measurements were made.
Time domain spectra from these measurements were found to be best fitted to the function
\begin{equation*}
A(t)=A_{1}{\rm e}^{-\lambda_{1} t} \textrm{cos}(\gamma_{\mu} B_{1} t)+A_{2}{\rm e}^{-\lambda_{2} t} \textrm{cos}(\gamma_{\mu} B_{2}t)+A_{\textrm{bg}}.
\end{equation*}
Figs~\ref{TFmuSR}(a-c) show the result of this fitting for GaV$_4$S$_8$ in an external field $\mu_{0}H=50$~mT [corresponding Fourier transformation spectra are shown in the SI\cite{SupMat}].
In the PM phase two internal fields are observed and most likely due to the existence of two inequivalent muon sites, consistent with the ZF results. The internal fields decrease as $T_{\mathrm{c}}$ is approached from above. The inset of Fig.~\ref{TFmuSR}(a) shows the corresponding muon Knight shifts
$K_{i}=(B_{i}-\mu_{0}H)/\mu_{0}H$
of both field components $B_{i}$ as a function of susceptibility $\chi$ above $T_{\textrm{c}}$.\cite{higemoto_muon_2010} The Knight shifts are negative with their magnitude increasing linearly with the susceptibility. This indicates significant hyperfine coupling, especially in the second muon site. In the magnetically ordered phase the second component of the internal field $B_{2}$ is not resolved. The relaxation rate $\lambda_{2}$, corresponding to the width of the distribution of the internal field $B_{2}$, shows a discontinuity at $T_{\mathrm{c}}$ and exhibits very large values in the magnetically ordered phases. The most likely explanation for this behaviour is a very large, fluctuating local field in the magnetically ordered phase at the second muon site. The relaxation rate is too large compared to the muon precession frequency, and thus the $B_2$ component is not observed. As a result, the effect of the fluctuations in $B_2$ result in a purely relaxing component with a large relaxation rate $\lambda_{2}$. 

Turning to the first component, a notable feature is an increase in the internal field component $B_{1}$ in the temperature range where the SkL phase is stabilised. This effect appears to provide a signature of the SkL phase in TF $\mu^{+}$SR measurement on this system. The relaxation rate $\lambda_{1}$ increases continuously upon cooling, and we do not observe a resolvable influence of the SkL on the relaxation rate. We thus conclude that the dynamics emerging with the SkL seen in the LF measurements do not have a strong effect along the direction of the applied magnetic field (as it is correlations along this direction that principally determine the relaxation in TF measurements). 

Results from fitting TF $\mu^{+}$SR measurements made on GaV$_4$Se$_8$ in $\mu_{0}$H$=150$~mT are presented in Figs~\ref{TFmuSR}(d-f). As for GaV$_4$S$_8$ two internal fields are observed at all applied magnetic fields, consistent with two inequivalent muon sites. No increase in $B_1$ can be observed in any part of the magnetically ordered phase, which was the signature of the SkL for GaV$_4$S$_8$.

\section{Muon site calculations}
\label{sec:musite}
\begin{figure}[t]
	\includegraphics[width=0.8\columnwidth]{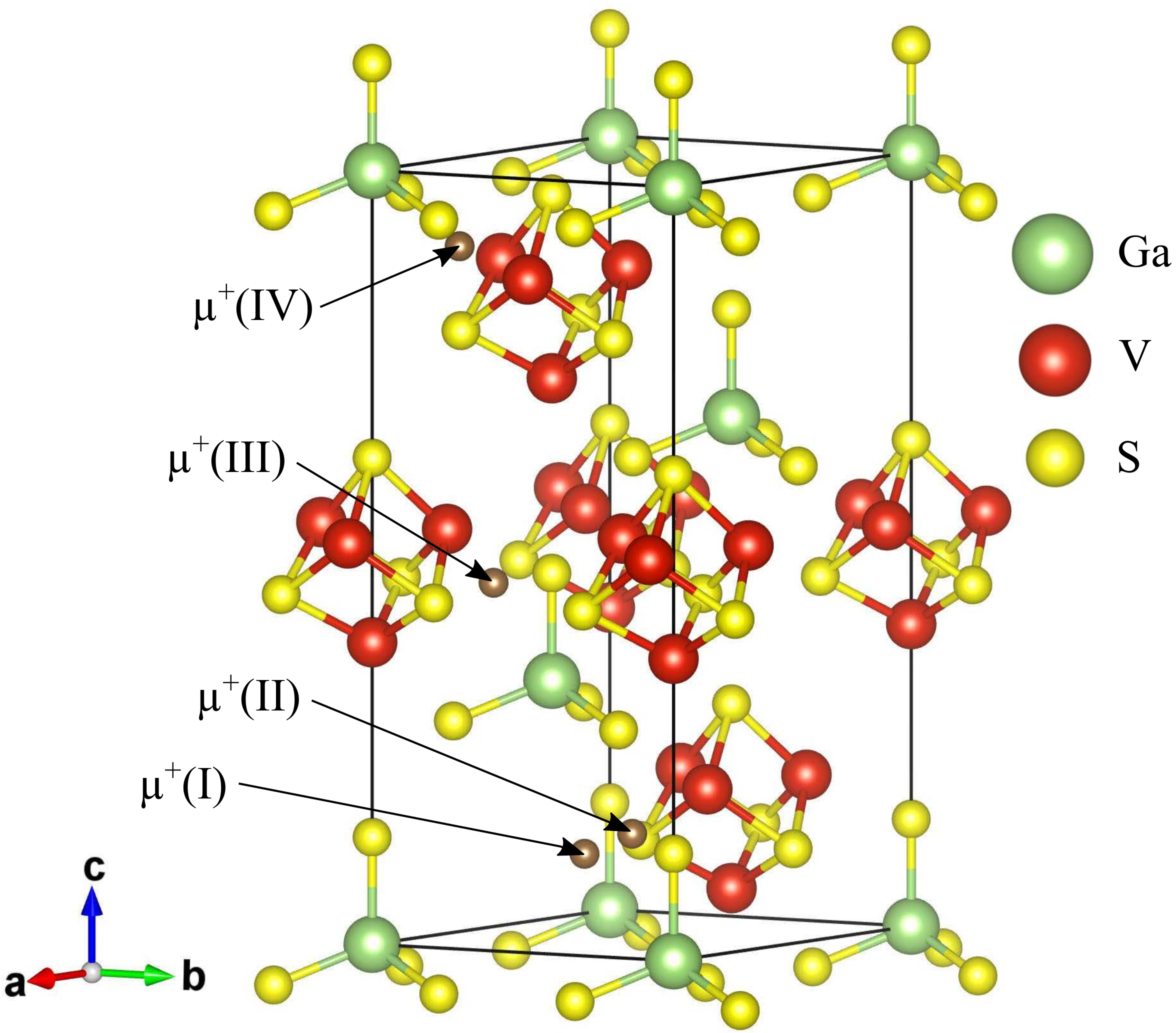}
	\caption{Four classes of muon stopping site determined for GaV$_4$S$_8$. The sites are numbered in order of increasing energy.}
	\label{GaVS_sites}
\end{figure}

\begin{figure}[t]
	\includegraphics[width=0.9\columnwidth, trim={0cm 0cm 0cm 0cm},clip]{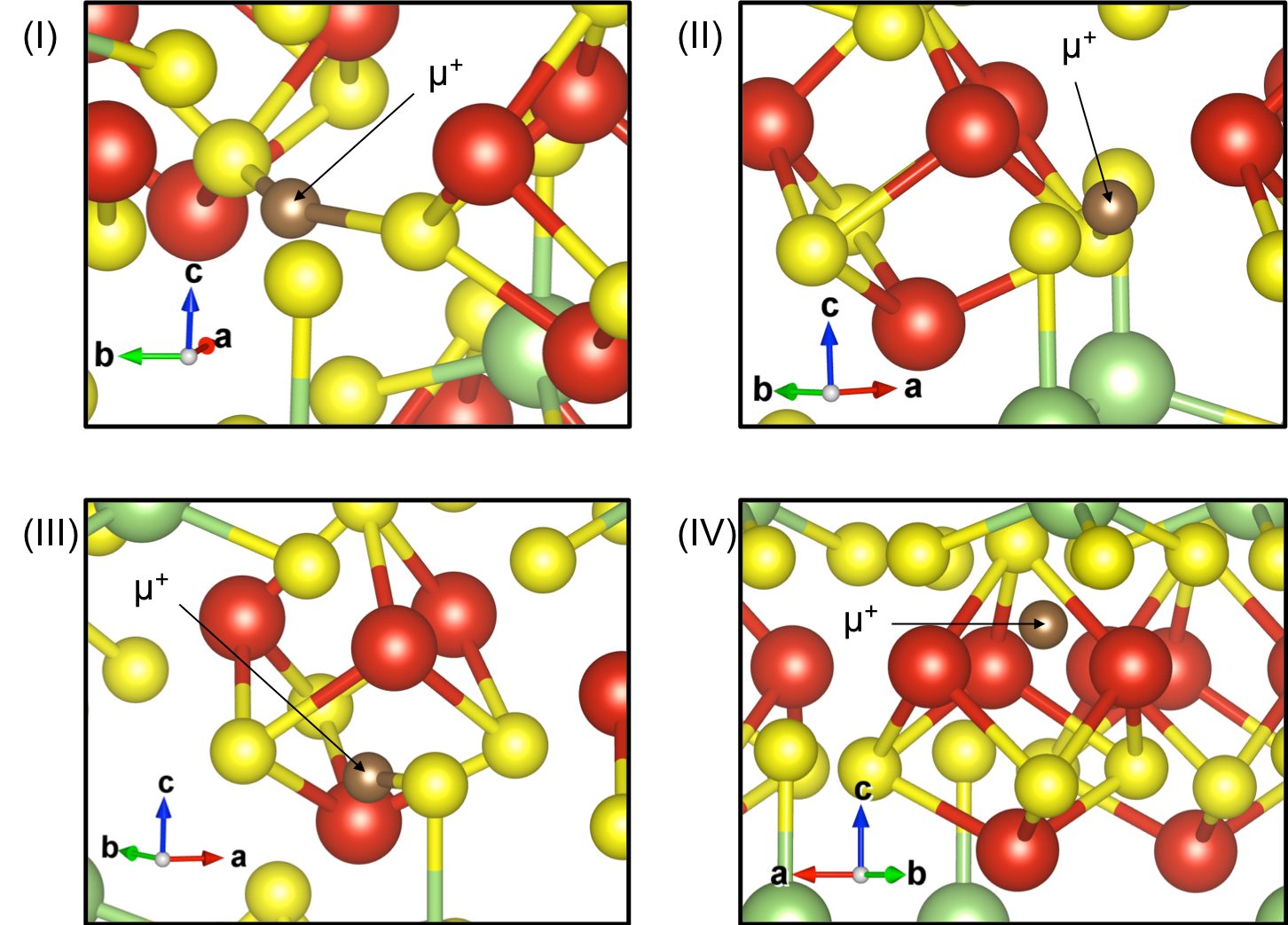}
	\caption{Local geometry around the muon for each of the four classes of muon stopping site in GaV$_{4}$S$_{8}$.}
	\label{GaVS_sites_local}
\end{figure}

To understand the details of the muon's interaction with the system, candidate muon sites were determined using DFT. Structural relaxations of a periodic supercell of GaV$_{4}$S$_{8}$ using DFT reveal four distinct candidate muon stopping sites [Fig.~\ref{GaVS_sites}], listed in Table~\ref{muon-sites}.
\begin{table}[b]
\centering
\setlength{\tabcolsep}{0.1cm}
\caption{\label{muon-sites}List of muon stopping sites with their relative energy $\Delta E$ and distance $d$ to the closest S or Se atom, respectively.}
\begin{tabular}{c c c @{\hspace{1.8cm}} c c c}
\hline \hline
& GaV$_4$S$_8$ & & & GaV$_4$Se$_8$ \\
Site & $\Delta E$ [eV] & $d$ [\AA] & Site & $\Delta E$ [eV] & $d$ [\AA] \\
\hline
I    	&	$0$		&	$1.4$	&	4	&	$0.381$	&	$1.6$	\\
II	&	$0.137$	&	$1.5$	&	3	&	$0.190$	&	$1.7$	\\
III	&	$0.288$	&	$1.4$	&	2	&	$0.145$	&	$1.7$	\\
IV	&	$0.293$	&	$2.0$		&	1	&	$0$		&	$2.2$	\\
\hline  \hline
\end{tabular}
\end{table}
Three of these (labeled I--III) involve the muon sitting close to a single S atom, which makes sense on chemical grounds, given the electronegativity of S. A fourth site (site IV) has the muon closer to V atoms and is the highest energy site. In the lowest energy site (site I) the muon sits between two S atoms, in the plane defined by three S atoms within V$_4$S$_4$ units [Fig.~\ref{GaVS_sites_local}(I)].  The two $\mu^+$--S distances are unequal (1.4 \AA~ and 2.0 \AA) with greater electron density found between the muon and the nearest S atom. (The S--$\mu^+$--S angle is $160^{\circ}$.)  This site is therefore best described in terms of the muon forming a $\mu^+$--S bond (rather than an S--$\mu^+$--S state by analogy to the commonly observed F--$\mu^+$--F complex \cite{FmuF}), though the presence of a second nearby S atom does seem to stabilize this geometry.
Two further sites involve the muon sitting close to a single S atom. In site II the muon sits along an edge of one of the V$_4$S$_4$ cubane-like units [Fig.~\ref{GaVS_sites_local}(II)].  This site is $0.137$~eV higher in energy than site I. The muon sits $1.5$~\AA~ from a S atom, which is similar to the shortest $\mu^+$-S distance for site~I.  
In site III [Fig.~\ref{GaVS_sites_local}(III)], the muon again sits $1.4$~\AA~ from an S atom, but this time the S atom belongs to a GaS$_4$ tetrahedron.  
Despite the similar coordination of the muon by S, this site is $0.288$~eV higher in energy than site I.
Unlike sites I--III, site IV does not involve the formation of a $\mu^+$--S bond. The muon sits above a face of one of the V$_4$S$_4$ cubane-like units, with the nearest S atom just over $2$~\AA\ away. 
The energy of this site is the highest, close to that of site 3: 0.293~eV higher in energy than site I.

The results of analogous calculations for GaV$_4$Se$_8$ are presented in Table~\ref{muon-sites}. Muon stopping sites (labeled $1$--$4$ in order of ascending energy) are similar to those calculated for GaV$_4$Se$_8$, with three of the four sites involving the muon sitting close to a Se atom (sites $2$ to $4$) and a site in which the muon sits above a face of a V$_4$Se$_4$ unit (site~1). However, the ordering of sites is inverted in this case. In particular, the cube face site (site~1), which corresponds to the highest energy stopping site for GaV$_4$S$_8$, is the lowest energy site for GaV$_4$Se$_8$. 
The three sites in which the muon sits near an Se atom have analogous sites in GaV$_4$S$_4$. Site~2 is $0.145$~eV higher in energy than the lowest energy site, and similar to site III in GaV$_4$S$_8$, with the muon bonded to a Se atom at the top of a GaSe$_4$ tetrahedron. 
Site~3 is $0.190$~eV higher in energy than the lowest energy site and is similar to site II in GaV$_{4}$S$_{8}$, but with a longer $\mu^+$--Se distance ($1.7$ \AA).  Site 4 (0.381~eV higher in energy than site 1) is similar to Site I in GaV$_{4}$S$_{8}$. 

We conclude that, although the sites for the two systems are similar owing to their similar structures, the energetic ordering of sites in the two cases is rather different. In particular, the lowest energy site, where we expect muons to stop, is different for the two materials despite their structural resemblance.  

\section{Discussion}
\label{sec:discussion}
ZF and TF $\mu^{+}$SR measurements suggest at least two magnetically inequivalent muon sites in GaV$_4$S$_{8}$ and GaV$_{4}$Se$_{8}$. Our DFT calculations reveal four cystallographically distinct candidate muon sites with different energies. They are however not necessarily all occupied, and in fact we expect the one with the lowest energy to be the most probable muon stopping site. The two magnetically inequivalent sites could thus be due to cystallographically different sites, or more likely structurally equivalent ones experiencing different internal fields due to the complex incommensurate spin textures in these systems. 

Nonetheless, the magnetic spin structures of these materials are based on an underlying length scale which is long compared to the size of the unit cell, making the details of exactly where the muon localizes within the unit cell relatively unimportant in comparing the different materials. 
We note that the location of the magnetic transitions, temperature and field-dependent details of the phase diagram, and collective properties of the system match with those measured using other techniques. Furthermore, we do not observe any significant structural distortions due to implanted muons in our DFT calculations. \cite{SupMat} We conclude, therefore, that the probe formed by the muon and any local distortion to the electronic structure appears to faithfully reflect the undistorted magnetism in this system.

ZF $\mu^{+}$SR on the S-containing material reveals an unusual increase of the total local magnetic field $B_{1}$ with temperature. This effect is not observed in the Se-containing analogue suggesting that the ground state magnetism in these systems is significantly different. Recent observations of the S-containing material suggest that, on cooling, the periodicity of the magnetic cycloid spin structure increases,  eventually transforming into a soliton lattice with periodically arranged domain walls in the FM phase.\cite{white_direct_2018,widmann_multiferroic_2016} 
This continuous change in magnetic structure might provide a mechanism for the unusual $T$-dependence of the local $B$-field. It would be expected to alter the sum of the dipolar fields at the muon sites, which could cause sufficient cancellation to cause the decrease in $B$ with decreasing temperature that we observe. Perhaps more significantly, as the cycloid unwinds and the structure becomes more ferromagnetic, there would be an increase in the size of the negative contribution from the demagnetizing field, further reducing $B$. It is notable that $B$ continues to decrease continuously through the 4~K, C-FM phase transition, marked in our data by a peak in the relaxation rate. This suggests that although some dynamic relaxation channels might freeze out at 4~K, the magnetic structure evolves continuously from the point of view of the muon ensemble. However, the need for a different phase offset in fitting the muon data in the FM region (often reflecting an incommensurate field distribution) is suggestive of a more complicated spin structure than simple ferromagnetism. 

The other possible contribution to the unusual temperature evolution of $B$ is the evolution of the hyperfine field at the muon site, arising from electronic spin density at the position of the implanted $\mu^{+}$.\cite{schenck_1999} The observation of significant Knight shifts in GaV$_4$S$_8$ but not GaV$_4$Se$_8$, along with the unusual temperature dependence of the energy gap in GaV$_{4}$S$_{8}$,\cite{sahoo_evidence_1993} means we might expect this contribution to be important. (We note, that a hyperfine field is likely required to reconcile the calculated dipole field at the muon site with that seen in ZF measurements.\cite{SupMat}) Specifically, in GaV$_4$S$_{8-y}$Se$_{y}$ compounds the magnetic units are metallically bonded V$_{4}$ tetrahedral clusters carrying an effective $S=1/2$ spin. Electronic conduction occurs by electron hopping between clusters and thus electronic properties depend strongly on details of the V$_{4}$ tetrahedra. While GaV$_{4}$Se$_{8}$ exhibits a constant energy gap over a wide temperature range, the energy gap of GaV$_{4}$S$_{8}$ decreases upon cooling and may even reach zero close to the magnetic phase transition. \cite{sahoo_evidence_1993} 
It may be that both the continuous change in magnetic structure and that of the energy gap reflect the same underlying $T$-dependent changes in the electronic structure.

A further notable feature of ZF measurements on GaV$_4$S$_{8}$ is the somewhat surprising difference in the temperature evolution of the relaxation rates $\lambda_{1}$ and $\lambda_{2}$. The relaxation rate $\lambda_{1}$ peaks at $T_{\mathrm{c}}$, as expected for the slowing of dynamics at a critical point, whereas $\lambda_{2}$ seems to reflect the behaviour of the magnitude of the local fields close the the transition, while also peaking at the FM to C transition. This is suggestive of the two muon sites coupling differently to the dynamics in the system and might suggest that the local environment at the two sites is more distinct than would typically be expected for two positions in the unit cell that are magnetically inequivalent. This could also explain why the two components in the TF spectra behave quite differently.

Based on our measurements on GaV$_4$S$_8$ we concluded that the SkL can be detected via a sizeable contribution to the relaxation rate in LF $\mu^{+}$SR measurements. An enhancement in $\lambda$ corresponds to an increase in the widths of the components of the field distribution, or an increase in the correlation time $\tau$ of the dynamic fluctuations (i.e.~a decrease in their fluctuation rate) perpendicular to the applied field direction, or to both. From our TF measurements we conclude that we do not observe a change in dynamics or field distribution specific to the SkL along the applied magnetic field direction. An increase in the relaxation rate in LF $\mu^{+}$SR measurements thus most likely reflects a contribution of the SkL to dynamics perpendicular to the applied field directions. As the dynamics probed by $\mu^{+}$SR can reach the GHz regime, our observation is attributable to the emergent excitation modes of individual skyrmions, that have been observed in the GHz regime, where clockwise, counterclockwise, and breathing modes occur in the skyrmion plane, but not perpendicular to it.\cite{ehlers_skyrmion_2016} 
It is notable that dynamics coinciding with the SkL occur over a broad spectral range, with
our AC susceptibility (on the kHz frequency scale) showing an increase in the imaginary component $\chi''$ in the SkL phases, indicating an increase in dissipation.\cite{SupMat} 
This increase is consistent with the increase in $\lambda$ seen in the muon results, since the fluctuation-dissipation theorem predicts that the spin correlation function $S(q,\omega=0)\propto T\lim_{\omega\rightarrow 0}\chi''(q, \omega)/\omega$ and we expect that the muon-spin relaxation
$\lambda \propto \sum_{q}A^{2}(q)S(q,0)$, where $A$ is the coupling of the muon to the spin system. 

The LF and TF $\mu^{+}$SR results for GaV$_4$Se$_8$ presented in Figs~\ref{LFmuSR}(h) and \ref{TFmuSR}(e) respectively, do not show the signature of a SkL phase for $\mu_{0}H=150$ mT. 
However, for $\mu_{0}H=95$~mT we observe a broadened peak in $\lambda$ down to $13.5$~K [Fig.~\ref{LFmuSR}(g)]. 
Following our interpretation of the LF results obtained on GaV$_4$S$_8$ we ascribe this increase in dynamics to the presence of the SkL phase. From AC susceptibility measurements [shown in Fig~\ref{ACsusc}(d)] we identify an increase in $\chi'$ around $13$~K and $100$~mT as the location of the SkL phase in agreement with $\mu^{+}$SR results. 
Previous reports of a SkL in single crystals of GaV$_4$Se$_8$ see this region of increased susceptibility as the transition from a C to a SkL phase that persists down to the lowest measured temperatures ($2$~K).\cite{fujima_thermodynamically_2017,bordacs_equilibrium_2017} 
Our muon results are therefore consistent with our AC susceptibility measurements, but do not match the phase diagram reported for GaV$_4$Se$_8$ single crystals from AC and DC susceptibility and from small angle neutron scattering (SANS).\cite{fujima_thermodynamically_2017, bordacs_equilibrium_2017}
The dynamic SkL response we observe in our GaV$_4$Se$_8$ sample is confined to a far smaller region of the field-temperature phase diagram.
The discrepancy could be due to the use of polycrystalline samples instead of single crystals, but we note that this does not lead to such a significant difference in the location and extent of the SkL phase in GaV$_4$S$_8$, where dynamics are observed over the temperature and field range where the SkL was seen in single crystals.\cite{kezsmarki_neel-type_2015,white_direct_2018}
It is thus plausible that the SkL region of GaV$_4$Se$_8$, is less extensive (at least in our polycrystalline sample) than previously suggested.

\section{Conclusions}
\label{sec:conclusions}
We have used $\mu^{+}$SR to investigate the skyrmion-lattice (SkL) phase in GaV$_4$S$_8$ and GaV$_4$Se$_8$.  While GaV$_4$S$_8$ and GaV$_4$Se$_8$ are structurally very similar and both exhibit a SkL phase, we have shown that their magnetic phase diagrams and ground states are significantly different and that the intermediate $y=2$ and 4 materials are glassy in their magnetic character. We have established the signature of the SkL in LF $\mu^{+}$SR  measurements, which has allowed us to observe characteristic dynamics on the MHz to GHz timescale. 
Our results suggest a phase diagram for polycrstalline GaV$_4$Se$_8$ in which the skyrmion phase appears substantially less extensive than reported in single crystal samples.

\begin{acknowledgments}
Part of this work was carried out at the Science and Technology Facilities Council (STFC) ISIS Facility, Rutherford Appleton Laboratory, UK and S$\mu$S, Paul Scherrer Institut, Switzerland. 
We gratefully acknowledge access to the MPMS in the Materials Characterisation Laboratory at ISIS.
We are grateful for the provision of beamtime and to A. Amato, H. Luetkens and J.S. Lord for experimental assistance.
DFT calculations were carried out using computing resources provided by STFC Scientific Computing Department's SCARF cluster and the Durham HPC Hamilton cluster.
We would like to thank M. N. Wilson and M. Gomil\v sek for fruitful discussion.
This work was supported by the EPSRC (EP/N032128/1 and EP/N024028/1).
Research data will be made available via Durham Collections.
\end{acknowledgments}

\end{document}